\documentclass{ifacconf}

\usepackage{style/mySetting}
\usepackage{graphicx}      
\usepackage{natbib}        

\begin{document}
\begin{frontmatter}

\title{
Small-Signal Stability Condition of Inverter-Integrated Power Systems:\\
Closed-Form Expression by Stationary Power Flow Variables
} 


\author[tech]{Taku Nishino} 
\author[tech]{Yoshiyuki Onishi} 
\author[tech]{Takayuki Ishizaki}

\address[tech]{Institute of Science Tokyo, Tokyo, Japan}



\begin{abstract} 
This paper shows that a necessary and sufficient condition for the small-signal stability of an inverter-integrated power system can be expressed in terms of semidefinite matrix inequalities determined only by the synchronous reactance of the components, the susceptance matrix of the transmission network, and the stationary values of the power flow distribution.
To derive the stability condition, we consider a class of grid-forming inverters corresponding to a singular perturbation of the synchronous generator.
The resulting matrix inequality condition, which has twice as many dimensions as the number of buses and is independent of the dynamics of the connected components, is expressed in terms of each component compensating in a decentralized manner for the loss of frequency synchronization caused by the reactive power consumption in the transmission network.
A simple numerical example using a 3-bus power system model shows that a grid-forming inverter load improves power system synchronization, while a grid-following inverter load disrupts it.
\end{abstract}
\begin{keyword} 
Power systems, Small-signal stability, Equilibrium-independent passivity. 
\end{keyword}

\end{frontmatter}

\section{Introduction}\label{sec:Introduction}

With the increase of DC power sources such as renewable energy and battery storage, there is a growing demand for stability analysis and stabilization control of power systems integrated with inverter resources \citep{SHAH2015ReviewPV}. 
In particular, research on grid-forming inverters that simulate the physical properties of frequency synchronization of conventional synchronous generators has attracted much attention. 
The analysis of power systems with such inverters is based on different assumptions than those of conventional power systems with large inertia synchronous generators, and general-purpose numerical analyses such as eigenvalue calculations \citep{sastry1980hierarchical} and time-response calculations \citep{stott1979power} dominate.

Against this background, this paper mathematically characterizes the small-signal stability of lossless power systems with synchronous generators and grid-forming inverters. 
The small-signal stability of power systems is equivalent to the local asymptotic stability of the equilibrium representing the power supply-demand balance. 
In particular, the following facts are proved in this paper.
\begin{itemize}
    \item[(i)]  A synchronous generator and a class of grid-forming inverters with identical synchronous reactances share the same decentralized stability condition.
    \item[(ii)]  The stability condition can be expressed in a closed form using only the solution of an algebraic equation for the stationary power flow distribution.
\end{itemize}

To clarify the contribution of this paper, we highlight its differences from existing small-signal stability analyses.
The references \citep{stegink2017unifying,DePersis2017,yang2019distributed} analyze stability using passivity-based approaches for power systems consisting of one-axis synchronous generator models, inverter-based resources, and a mix of both, respectively. 
Similarly, this paper also adopts a passivity-based approach to prove the sufficiency of small-signal stability.
However, none of the above references makes a clear distinction between the bus voltage and the internal voltage of the component connected to each bus.
As a result, the mathematical model of the synchronous generator assumes dynamics that are partially inconsistent with the physics of the AC circuit. 
In particular, while the differential equation of the generator includes a certain reactance constant, the reactance in the equation for the connection between the generator and the bus, which should be assumed to have the same value, is implicitly assumed to be infinitesimal. 
Therefore, it is not possible to derive from the assumptions of these references the equivalence with respect to the synchronous reactance that this paper clarifies as fact (i). 
In addition, the derived stability condition cannot be expressed in a closed form of the stationary power flow distribution, as in fact (ii).

It should be emphasized that this paper uses a standard synchronous generator model consistent with a proper AC circuit physics following \citep{sauer2017power}, and employs a more detailed two-axis model. 
Although omitted for space reasons, the same conclusions can be drawn using a more detailed Park generator model.

This paper is organized as follows.
Section \ref{sec:psys_model} defines the power system model to be analized. 
Section \ref{sec:main} presents a necessary and sufficient condition for the small-signal stability and its proof.  
Section \ref{sec:numerical_analysis} shows a numerical example using a 3-bus power system model to demonstrate that a grid-forming inverter load improves power system synchronization, while a grid-following inverter load disrupts it.
Section \ref{sec:conc} concludes the paper.

\section{Power System Model}\label{sec:psys_model}

\subsection{Transmission Network Model}

Consider a transmission network with $N$ buses. 
Without loss of generality, we assume that each bus is connected to a generator, load, or other device. 
Let $b_{ij} > 0$ be the susceptance of the transmission line connecting Buses $i$ and $j$. 
The susceptance matrix $B \in \mathbb{R}^{N \times N}$ is given by
\begin{equation}\label{eq:B}
    B_{ij} = \left\{\begin{array}{cl}
        b_{ij}, &\quad i \neq j\\
        \displaystyle -\sum_{j=1}^{N}b_{ij},  &\quad i=j.
    \end{array}\right.
\end{equation}
This is negative semidefinite because $-B$ is the weighted Laplacian of an undirected graph.

Let $\bm{I}_i \in \mathbb{C}$ and $\bm{V}_i \in \mathbb{C}$ be the current and voltage phasors at Bus $i$, respectively. 
Assuming zero line conductances, i.e., lossless, the power balance at each bus is given as
\begin{equation}\label{eq:PQ_def}
    \simode{
        P_i & = \sum_{j=1}^N B_{ij} V_i V_j \sin (\theta_i -\theta_j) \\
        Q_i & = \sum_{j=1}^N - B_{ij} V_i V_j \cos (\theta_i -\theta_j)
    }
\quad i\in \N
\end{equation}
where $\N$ is the label set of buses, $V_i$ and $\theta_i$ are the magnitude and phase of $\bm{V}_i$.
Furthermore, $P_i$ and $Q_i$ are the active and reactive powers defined as the real and imaginary parts of $\bm{V}_i \overline{\bm{I}_i}$, respectively.  
The power balance equation in \eqref{eq:PQ_def} can be simply written as
\begin{equation}\label{eq:PQ_def2}
0= g(\theta,V,P,Q).
\end{equation}

Since only the phase difference of voltage phasors matters, we introduce the following notion.

\begin{dfn}
Consider a stationary power flow distribution $(\theta^{\star}, V^{\star},P^{\star},Q^{\star})$ such that
\[
0= g(\theta^{\star},V^{\star},P^{\star},Q^{\star}).
\]
The set of equivalent stationary power flow distributions obtained by shifting voltage phases as
\[
(\theta_{\rm e}^{\star},V^{\star},P^{\star},Q^{\star}) = 
\bigl\{
(\theta^{\star}+c\mathds{1}_N, V^{\star},P^{\star},Q^{\star}): 
c\in \mathbb{R}
\bigr\}
\]
is called a \textbf{stationary power flow distribution set}.
\end{dfn}

For simplicity, we use the notation of
\[
\varpi^{\star}:=(\theta^{\star},V^{\star},P^{\star},Q^{\star})
,\quad
\varpi_{\rm e}^{\star}:=(\theta_{\rm e}^{\star},V^{\star},P^{\star},Q^{\star}).
\]
In particular, we simply write $f(\varpi_{\rm e}^{\star})$ if a function $f(\varpi^{\star})$ is invariant for any $\varpi^{\star}$ in $\varpi_{\rm e}^{\star}$.

\subsection{Synchronous Generator Model}\label{subsec:modeling_generator}

In this paper, we consider the two-axis synchronous generator model \citep{sauer2017power}, which represents the electromechanical dynamics along two orthogonal axes called the d-axis and the q-axis. 
The dynamics of the two-axis generator model is given by
\begin{subequations}\label{eq:2axis_model}
\begin{equation}\label{eq:2axis_state_DAE}
\simode{
\dot{\delta}_i &= \omega_0 \omega_i \\
M_i  \dot{ \omega}_i &= -D_i \omega_i - P_i + P^{\star}_{\mathrm{m}i} \\
\tau_{\mathrm{d}i} \dot{E}_{\mathrm{q}i} &= - E_{\mathrm{q}i}
 - (X_{\mathrm{d}i} - X'_{\mathrm{d}i}) I_{\mathrm{d}i} + V^{\star}_{\mathrm{fd}i} \\
\tau_{\mathrm{q}i} \dot{E}_{\mathrm{d}i} &= - E_{\mathrm{d}i}  + (X_{\mathrm{q}i} - X'_{\mathrm{q}i}) I_{\mathrm{q}i}
}
\end{equation}
where $E_{\mathrm{q}i} \in \mathbb{R}$ is the q-axis internal voltage for the field winding, 
$E_{\mathrm{d}i} \in \mathbb{R}$ is the d-axis internal voltage for the damper winding, 
$\delta_i \in \mathbb{R}$ is the rotor angle relative to a reference frame rotating at the nominal angular frequency $\omega_0 \in \mathbb{R}$, and 
$\omega_i \in \mathbb{R}$ is the angular frequency deviation from $\omega_0$. 
For the positive constants,
$M_i$ represents the inertia constant,
$D_i$ the damping coefficient,
$\tau_{\mathrm{d}i}$, $\tau_{\mathrm{q}i}$ the time constants of the internal voltages,
$X_{\mathrm{d}i}$, $X_{\mathrm{q}i}$ the synchronous reactances
$X'_{\mathrm{d}i}$, $X'_{\mathrm{q}i}$, the transient reactances,
$P^{\star}_{\mathrm{m}i}$ the mechanical power, and
$V_{\mathrm{fd}i}^{\star}$ the field voltage.

The connection between the generator and bus is given by
\begin{equation}\label{eq:2axis_output_I}
\textstyle
    I_{\mathrm{d}i} = \frac{1}{X'_{\mathrm{d}i}} ( E_{\mathrm{q}i} - V_{\mathrm{q}i} ) 
    ,\quad
    I_{\mathrm{q}i} = \frac{1}{X'_{\mathrm{q}i}} ( V_{\mathrm{d}i} - E_{\mathrm{d}i} )
\end{equation}
where $V_{\mathrm{q}i}$ and $V_{\mathrm{d}i}$ are defined as
\begin{equation*}\label{eq:Vq_Vd_def}
    V_{\mathrm{q}i}  = V_i \cos (\delta_i - \theta_i )
    ,\quad
    V_{\mathrm{d}i}  = V_i \sin (\delta_i - \theta_i ). 
\end{equation*}
Note that \eqref{eq:2axis_output_I} can be expressed in terms of the active power $P_i$ and reactive power $Q_i$ supplied by the generator to the bus as
\begin{equation}\label{eq:2axis_output_PQ}
\simode{
    P_i &= \textstyle \frac{E_{\mathrm{q}i}}{X'_{\mathrm{d}i}} V_{\mathrm{d}i} - \frac{E_{\mathrm{d}i}}{X'_{\mathrm{q}i}} V_{\mathrm{q}i} + \left( \frac{1}{X'_{\mathrm{q}i}}- \frac{1}{X'_{\mathrm{d}i}}\right) V_{\mathrm{d}i} V_{\mathrm{q}i}\\
    Q_i &= \textstyle \frac{E_{\mathrm{q}i}}{X'_{\mathrm{d}i}} V_{\mathrm{q}i} + \frac{E_{\mathrm{d}i}}{X'_{\mathrm{q}i}} V_{\mathrm{d}i} - \left( \frac{ V^2_{\mathrm{d}i}}{X'_{\mathrm{q}i}} + \frac{ V^2_{\mathrm{q}i}}{X'_{\mathrm{d}i}}\right).\\
}
\end{equation}
\end{subequations}
The synchronous and transient reactances generally satisfy 
\begin{equation}\label{eq:reactance}
    X'_{\mathrm{d}i} < X_{\mathrm{d}i}
    ,\quad
    X'_{\mathrm{q}i} < X_{\mathrm{q}i}.
\end{equation}

In the following, we represent the bus variables as
\[
v_i:=(\theta_i,V_i)
,\quad
w_i:=(P_i,Q_i).
\]
Choosing them as the input and output for the connection to the main grid, the dynamics in \eqref{eq:2axis_model} can be formally expressed as
\begin{equation}\label{eq:formal_model}
\simode{
\dot{x}_i &= f_i (x_i, v_i; u_i^{\star}) \\
w_i &= h_i (x_i, v_i; u_i^{\star})
}
\end{equation}
where $x_i$ is the state variable, and $u_i^{\star}$ is a constant input representing the mechanical power and field voltage.  

It is known that the stationary state $x_i^{\star}$ and the constant input $u_i^{\star}$ that achieve the desired stationary power flow $\varpi_{i}^{\star}$ at Bus $i$ in equilibrium such that
\[
0 = f_i (x_i^{\star}, v_i^{\star}; u_i^{\star}) 
,\quad
w_i^{\star} = h_i (x_i^{\star}; v_i^{\star}, u_i^{\star})
\]
are uniquely determined. 
In particular, the phase difference $\delta^{\star}_i - \theta^{\star}_i$ between the generator and bus at the stationary power flow $\varpi_{{\rm e}i}^{\star}$ is uniquely determined as
\begin{equation}\label{eq:def_phi}
\phi_i(\varrho_i^{\star}) := \atan \biggl( \frac{P_i^{\star}}{Q_i^{\star} + \frac{(V_i^{\star})^2}{X_{{\rm q}i}}} \biggr)
\end{equation}
where the set of the stationary voltage, and active and reactive powers is denoted by
\[
\varrho_i^{\star}:=(V_i^{\star},P_i^{\star},Q_i^{\star}).
\]
From this fact, the value of $u_i^{\star}$ is found as
\begin{equation}\label{eq:inputst}
\spliteq{
P_{{\rm m}i}^{\star} & = P^{\star}_i, \\
V_{{\rm fd}i}^{\star} & = \! \tfrac{ X_{{\rm d}i}P_i^{\star} }{V_i^{\star}} \sin \phi_i(\varrho_i^{\star})  \!+ \!
\left( \! \tfrac{ X_{{\rm d}i}Q_i^{\star} }{V_i^{\star}} \!+ \!V_i^{\star} \! \right) \cos \phi_i(\varrho_i^{\star}).
}
\end{equation}
Note that these are invariant with respect to the stationary power flow set $\varpi_{{\rm e}i}^{\star}$. 
In the following, we assume that the mechanical power and field voltage are fixed to the values in \eqref{eq:inputst} for the considered stationary power flow set.

\subsection{Grid-Forming Inverter Models}\label{subsec:modeling_inverter}

We consider two inverter-based resource models: the virtual synchronous generator (VSG) and the frequency droop control (FDC). 
These models represent devices that consume or supply power according to references.
We note that both are classified in the class of basic grid-forming inverter models that regulate the internal voltage to act as a voltage source; see 
\citep{schiffer2016survey} for details.

\subsubsection{VSG Model}\label{subsubsec:modeling_VSG}
The VSG mimics the classical synchronous generator model, derived from the two-axis model by assuming fast electromagnetic dynamics.  
In particular, assuming that the internal voltages are governed by the algebraic equations
\[
\simode{
0 &= - E_{\mathrm{q}i}
 - (X_{\mathrm{d}i} - X'_{\mathrm{d}i}) I_{\mathrm{d}i} + V^{\star}_{\mathrm{fd}i} \\
0 &= - E_{\mathrm{d}i}  + (X_{\mathrm{q}i} - X'_{\mathrm{q}i}) I_{\mathrm{q}i},
}
\]
we have the simple swing equation
\begin{subequations}\label{eq:classical_model}
\begin{equation}\label{eq:classical_state_DAE}
\simode{
\dot{\delta}_i &= \omega_0 \omega_i \\
M_i \dot{ \omega}_i &= -D_i \omega_i - P_i + P^{\star}_{\mathrm{m}i} .
}
\end{equation}
The connection to the grid is written as
\begin{equation}\label{eq:classical_output_I}
\textstyle
    I_{\mathrm{d}i} = \frac{1}{X_{\mathrm{d}i}} ( V^{\star}_{\mathrm{fd}i} - V_{\mathrm{q}i} ) 
    ,\quad
    I_{\mathrm{q}i} = \frac{1}{X_{\mathrm{q}i}} V_{\mathrm{d}i},
\end{equation}
or equivalently
\begin{equation}\label{eq:classical_output_PQ}
\simode{
    P_i &= \textstyle \frac{V^{\star}_{\mathrm{fd}i}}{X_{\mathrm{d}i}} V_{\mathrm{d}i} + \left( \frac{1}{X_{\mathrm{q}i}}- \frac{1}{X_{\mathrm{d}i}}\right) V_{\mathrm{d}i} V_{\mathrm{q}i}\\
    Q_i &= \textstyle \frac{V^{\star}_{\mathrm{fd}i}}{X_{\mathrm{d}i}} V_{\mathrm{q}i} - \left( \frac{ V^2_{\mathrm{d}i}}{X_{\mathrm{q}i}} + \frac{ V^2_{\mathrm{q}i}}{X_{\mathrm{d}i}}\right).
}
\end{equation}
\end{subequations}
Note that $P^{\star}_{{\rm m}i}$ and $V^{\star}_{{\rm fd}i}$ should be set as in \eqref{eq:inputst} to achieve the desired stationary power flow.

\subsubsection{FDC Model}\label{subsubsec:modeling_FDC}
The FDC model is derived assuming small inertia in the VSG model. 
The state equation becomes
\begin{subequations}
\begin{equation}\label{eq:droop_state_DAE}
D_i \dot{\delta}_i = \omega_0 (- P_i + P^{\star}_{\mathrm{m}i}).
\end{equation}
\end{subequations}
Similar to the VSG model, the connection equation is given by \eqref{eq:classical_output_I} or \eqref{eq:classical_output_PQ}, and $P^{\star}_{{\rm m}i}$ and $V^{\star}_{{\rm fd}i}$ are set by \eqref{eq:inputst}.

\subsection{Entire Power System Model}
The entire power system model is obtained as a nonlinear differential-algebraic equation system
\begin{equation}\label{eq:formal_DAE}
\simode{
\dot{x} &= f (x,v;u^{\star}) \\
w &= h(x,v;u^{\star}) \\
0 &= g(v,w)
}
\end{equation}
where a unique stationary state $x^{\star}$ is ensured for each $\varpi^{\star}$.
In the following, we denote the equilibrium set as $(x_{\rm e}^{\star},\varpi_{\rm e}^{\star})$ where $x_{\rm e}^{\star}$ denotes the set of $x^{\star}$ compatible with $\varpi_{\rm e}^{\star}$.

\section{Main Result}\label{sec:main}

\subsection{Characterization of Small-Signal Stability}
This section presents a necessary and sufficient condition for the small-signal stability of the power system model given in Section \ref{sec:psys_model}. 
In fact, the derived condition is independent of the details of the component models connected to each bus, and is expressed in a closed form using only the stationary power flow distribution.

The small-signal stability is formally defined as follows.

\begin{dfn}
A stationary power flow distribution set $\varpi_{\rm e}^{\star}$ is said to be \textbf{small-signal stable} if the equilibrium set $(x_{\rm e}^{\star},\varpi_{\rm e}^{\star})$ of the power system model in \eqref{eq:formal_DAE} is locally asymptotically stable.
\end{dfn}

In the following, the d-axis and q-axis synchronous reactances are represented together as
\[
X_i:=(X_{\mathrm{d}i}, X_{\mathrm{q}i}).
\]
Then, the main result of this paper is stated as follows.

\begin{thm}\label{thm:main}\normalfont
A stationary power flow distribution set $\varpi_{\rm e}^{\star}$ is small-signal stable if and only if
\begin{subequations}\label{eq:cnd_iff}
\begin{align} \label{eq:cnd_gam}
&\sfdiag \bigl( 
\gamma_i (\varrho_i^{\star};X_i)\bigr) \succ 0, \\
&\sfdiag \bigl( \varGamma_{i}(\varrho_i^{\star};X_i ) \bigr) 
+L(\theta_{\rm e}^{\star}, V^{\star}; B) 
\succeq 0
\label{eq:cnd_mats}
\end{align}
\end{subequations}
where the scalar $\gamma_i$ is defined by
\begin{equation}\label{eq:def_gam}
\gamma_i (\varrho_i; X_i) := 
Q_i +
\frac{ V_i^2 \cos^2 \phi_i (\varrho_i) }{X_{\mathrm{q}i}} +
\frac{ V_i^2 \sin^2 \phi_i (\varrho_i)}{X_{\mathrm{d}i}}
\end{equation}
with $\phi_i$ in \eqref{eq:def_phi}, the submatrix $\varGamma_{i}$ and the submatrix $L_{ij}$ of 
\[
L(\theta_{\rm e}^{\star}, V^{\star}; B) := 
\sum_{i=1}^N \sum_{j=1}^N \left\{
e_i e_j^{\sf T} \otimes L_{ij}(\theta_{\rm e}^{\star}, V^{\star}; B)
\right\}
\]
are defined as in \eqref{eq:def_mats}, and $e_i$ denotes the $i$th canonical unit vector of dimension $N$.
\end{thm}

\newcounter{MYtempeqncnt}
\begin{figure*}[!t]
\normalsize
\setcounter{MYtempeqncnt}{\value{equation}}
\setcounter{equation}{14}
\begin{equation}\label{eq:def_mats}
\spliteq{
\varGamma_i(\varrho_i;X_i)
 &:= 
\mat{
0 & 0 \\ 0 &
\frac{
\frac{ V_i^4 }{ X_{\mathrm{q}i} X_{\mathrm{d}i} } 
- P_i^2
+
\Bigl(
\frac{ V_i^2 \cos^2 \phi_i(\varrho_i) }{ X_{\mathrm{d}i} }  +
\frac{ V_i^2 \sin^2 \phi_i(\varrho_i) }{ X_{\mathrm{q}i} }
\Bigr) Q_i  -
2 \left( \frac{ 1 }{ X_{\mathrm{q}i}} - \frac{1}{ X_{\mathrm{d}i} } \right)  P_i
V_i^2 \cos \phi_i(\varrho_i) \sin \phi_i(\varrho_i)
}{
V_i^2 
\gamma_i (\varrho_i; X_i)
}
} \\
L_{ij}(\theta,V;B) &:= \left\{ \begin{array}{cl}
\mat{
\sum_{j\neq i}^N\limits B_{ij} V_iV_j \cos (\theta_i -\theta_j) &
\sum_{j=1 }^N\limits B_{ij} V_j \sin (\theta_i -\theta_j) \\ 
\sum_{j=1 }^N\limits B_{ij} V_j \sin (\theta_i -\theta_j) &
-B_{ii} 
}, &\quad i=j \vspace{2mm}\\
\mat{
-B_{ij} V_iV_j \cos (\theta_i -\theta_j) &
 B_{ij} V_i \sin (\theta_i -\theta_j) \\ 
- B_{ij} V_j \sin (\theta_i -\theta_j) &
-B_{ij} \cos (\theta_i -\theta_j)
}, &\quad i\neq j 
\end{array}
\right. \\
}
\hspace{-10mm}
\end{equation}
\setcounter{equation}{\value{MYtempeqncnt}}
\hrulefill
\vspace*{4pt}
\end{figure*}

\addtocounter{equation}{1}

The stability condition in Theorem \ref{thm:main} consists of two parts: 
a local condition for each device and a global condition for the entire transmission network. 
In particular, \eqref{eq:cnd_gam} and the first term of \eqref{eq:cnd_mats} are local conditions for each component, while the second term of \eqref{eq:cnd_mats} is a global condition for the transmission network.
Surprisingly, \eqref{eq:def_mats} does not contain a stationary state of the ``dynamics."

The matrix $L$ in \eqref{eq:cnd_mats} is determined solely by the susceptance matrix of the transmission network and the stationary values of the voltage phasors. 
In particular, the eigenvalues of $L$ become more negative as the phase differences between the bus voltage phasors increase. 
From this perspective, $L$ characterizes the stability of the transmission network with respect to the voltage distribution. 
Note that $L$ is not generally positive semidefinite.

On the other hand, $\varGamma_i$ is determined only by the synchronous reactance of each component and the stationary power flow distribution, and becomes positive semidefinite under appropriate conditions. 
In particular, the eigenvalues of $\varGamma_i$ become more positive as each component contributes more strongly to the frequency synchronization of the entire power system.

\subsection{Proof of Sufficiency} \label{subsec:sufficient_cond}
This section proves that \eqref{eq:cnd_iff} is a sufficient condition for the small-signal stability, using the concept of equilibrium-independent passivity \citep{hines2011equilibrium, simpson2018equilibrium}. 
We analyze the input-output characteristics of each bus, considering it as a subsystem.

\subsubsection{Storage Function Candidates}
We consider Bus $i$ as a subsystem with the input $(-P_i, -\tfrac{Q_i}{V_i})$ and output $(\dot{\theta}_i,\dot{V}_i)$. 
Let $z_i$ be the vector stacking the internal state of the component and voltage phasor, denoted by $x_i$ and $v_i$, respectively. 
We define energy functions $U_i(z_i)$ as
\begin{equation} \label{eq:2axis_potential}
\spliteq{
U_i (z_i)
    =
    \tfrac{\omega_0 M_i \omega_i^2}{2} &+ 
    \tfrac{E_{\mathrm{q}i}^2}{ 2(X_{\mathrm{d}i} - X'_{\mathrm{d}i})}
    +
    \tfrac{E_{\mathrm{d}i}^2}{ 2(X_{\mathrm{q}i} - X'_{\mathrm{q}i})}\\
    &+
    \tfrac{( V_{\mathrm{d}i} - E_{\mathrm{d}i} )^2 }{2 X'_{\mathrm{q}i}}
    +
    \tfrac{(E_{\mathrm{q}i} - V_{\mathrm{q}i} )^2 }{2 X'_{\mathrm{d}i}}
}
\end{equation}
for the two-axis generator model.
Furthermore, we define
\begin{equation} \label{eq:classical_potential}
U_i (z_i)
    =
    \tfrac{\omega_0 M_i \omega_i^2}{2}
    +
    \tfrac{V^2_{\mathrm{d}i}}{2 X_{\mathrm{q}i}}
    +
    \tfrac{( V^{\star}_{\mathrm{fd}i} - V_{\mathrm{q}i} )^2 }{2 X_{\mathrm{d}i}}
\end{equation}
for the VSG model, and
\begin{equation} \label{eq:droop_potential}
U_i (z_i)
    =
    \tfrac{V^2_{\mathrm{d}i}}{2 X_{\mathrm{q}i}}
    +
    \tfrac{( V^{\star}_{\mathrm{fd}i} - V_{\mathrm{q}i} )^2 }{2 X_{\mathrm{d}i}}
\end{equation}
for the FDC model.

For each model, we use the Bregman distance 
\[
W_i(z_i):=U_i(z_i)-U_i(z_{{\rm e}i}^{\star}) - \nabla U_i(z_{{\rm e}i}^{\star}) (z_i-z_{{\rm e}i}^{\star})
\]
as a storage function candidate.
On the other hand, for the transmission network, with the input $(\dot{\theta},\dot{V})$ and output $(P, \tfrac{Q}{V})$, we define the energy function as
\[
U_0 (v) = -\frac{1}{2} \sum_{i=1}^N \sum_{j=1}^N 
B_{ij} V_i V_j \cos (\theta_i - \theta_j)
\]
and the storage function candidate as
\[
W_0(v):=U_0(v)-U_0(v_{\rm e}^{\star}) - \nabla U_0(v_{\rm e}^{\star}) (v-v_{\rm e}^{\star}).
\]

\subsubsection{Lyapunov Inequality}
Consider the candidate of the Lyapunov function
\[
W(x,v) := W_0(v) + \sum_{i=1}^N W_i(x_i,v_i).
\]
Then, its time derivative is obtained as
\[
\spliteq{
\dot{W}(x,v) = & - \sum_{i\in \mathcal{N} } \tfrac{D_i \dot{\delta}_i^2 }{\omega_0}  
- \sum_{i\in \mathcal{M}} 
\left(
\tfrac{\tau_{\mathrm{d}i} \dot{E}_{\mathrm{q}i}^2}{X_{\mathrm{d}i} - X'_{\mathrm{d}i}} 
+ \tfrac{\tau_{\mathrm{q}i} \dot{E}_{\mathrm{d}i}^2 }{X_{\mathrm{q}i} - X'_{\mathrm{q}i}}  
\right) 
}
\]
where $\mathcal{M}$ is the set of buses with the two-axis model. 
Thus, as long as $W(x,v)$ is positive semidefinite around the equilibrium, all $\dot{\delta}_i$, $\dot{E}_{{\rm q}i}$, and $\dot{E}_{{\rm d}i}$ converge to 0, implying that the stationary state is attained.

\subsubsection{Positive Semidefiniteness of Storage Functions}\label{sec:Hposisemi}
To ensure that the Lyapunov function is positive semidefinite, we analyze the Hessian of the energy functions.
First, we consider the two-axis generator model. 
Denote $E_{{\rm q}i}$ and $E_{{\rm d}i}$ together as $E_i$.
Then, the Hessian is obtained as
\[
\nabla^2 U_i (x_i, v_i ) =
\underbrace{
\mat{
\hat{H}_{\omega_i\omega_i} & 0 & 0 & 0\\
0 &\hat{H}'_{\delta_i \delta_i} & \hat{H}'_{\delta_i E_i} & \hat{H}'_{\delta_i v_i} \\
0 &* & \hat{H}'_{E_i E_i}  &\hat{H}'_{E_i v_i}  \\
0 &* & * & \hat{H}'_{v_i v_i}
}
}_{\hat{H}'}
\]
From \eqref{eq:reactance}, we see that $\hat{H}'_{E_i E_i}$ is positive definite. 
Computing the Schur complement $\hat{H}'/\hat{H}'_{E_i E_i}$, we obtain
\[
\nabla^2 U_i (x_i,v_i ) =
\mat{
\hat{H}_{\omega_i\omega_i} & 0 & 0\\
0 & \hat{H}_{\delta_i \delta_i}  & \hat{H}_{\delta_i v_i} \\
0 & * & \hat{H}_{v_i v_i} 
}
\]
where the submatrices are obtained as
\[
\spliteq{
\hat{H}_{\delta_i \delta_i} &= \textstyle
\frac{{V^2_{\mathrm{q}i}}}{X_{\mathrm{q}i}}
+
\frac{{V^2_{\mathrm{d}i}}}{X_{\mathrm{d}i}}
+
Q_i, \\
\hat{H}_{\delta_i v_i} &= 
\mat{
- \hat{H}_{\delta_i \delta_i} &
\tfrac{1}{ V_i } \Big\{ P_i + 
\big( \tfrac{1}{X_{\mathrm{q}i}} - \tfrac{1}{X_{\mathrm{d}i}} \big) V_{\mathrm{d}i} V_{\mathrm{q}i} \Big\}
}, \\
\hat{H}_{v_i v_i} & =\mat{
\hat{H}_{\delta_i \delta_i}
&
-\tfrac{1}{ V_i } \Big\{ P_i 
+ \big( \tfrac{1}{X_{\mathrm{q}i}} - \tfrac{1}{X_{\mathrm{d}i}} \big) V_{\mathrm{d}i} V_{\mathrm{q}i} \Big\} \\
* &
\tfrac{1}{ V_i^2 } \Big( 
\tfrac{{V^2_{\mathrm{d}i}}}{X_{\mathrm{q}i}} + 
\tfrac{{V^2_{\mathrm{q}i}}}{X_{\mathrm{d}i}} 
\Big)
}.
}
\]
In fact, the resultant Schur complement coincides with the Hessian of the energy function for the VSG model. 
In addition, since $\hat{H}_{\omega_i\omega_i}$ is positive definite, its positive semidefiniteness is equivalent to that of
\begin{equation}\label{eq:redHess}
\nabla^2 U_i (x_i,v_i ) =
\mat{
\hat{H}_{\delta_i \delta_i}  & \hat{H}_{\delta_i v_i} \\
* & \hat{H}_{v_i v_i} 
},
\end{equation}
which is the Hessian of the energy function for the FDC model.
This implies that the positive semidefiniteness of the storage function is equivalent for all models.

For the transmission network, the Hessian is 
\[
\nabla^2 U_0 (v ) =\mat{
\check{H}_{v_1 v_1} &  \cdots & \check{H}_{v_1 v_N} \\
\vdots & \ddots & \vdots \\
* &  \cdots & \check{H}_{v_N v_N} 
}
\]
where the submatrices are found to be
\begin{equation}\label{eq:net_Hess}
\check{H}_{v_i v_i} = L_{ii}(v;B)
,\quad
\check{H}_{v_i v_j} = L_{ij}(v;B)
\end{equation}
with $L_{ij}$ defined in \eqref{eq:def_mats}.

\subsubsection{Positive Semidefiniteness of Lyapunov Function}
To prove the positive semidefiniteness of the Lyapunov function, we analyze the Hessian of the total energy function
\[
U(x,v) = U_0(v) + \sum_{i=1}^N U_i(x_i,v_i).
\]
Since the internal state $x_i$ only affects the energy function at Bus $i$, we can compute the Schur complements with respect to the variables other than $\delta_i$ independently. 
Therefore, it suffices to consider \eqref{eq:redHess} as the Hessian of the bus energy function.  
This leads to
\[
\nabla^2 U(x,v)=\sum_{i=1}^N \sum_{j=1}^N \Bigl( e_i e_j^{\sf T} \otimes H_{ij} \Bigr)
\]
where the submatrices are defined as
\[
H_{ij} := \left\{ \begin{array}{lc}
\mat{ \hat{H}_{\delta_i \delta_i} & \hat{H}_{\delta_i v_i} \\
* & \hat{H}_{v_i v_i} + \check{H}_{v_i v_i} \\
}, & \quad i = j \\
\mat{ 0 & 0 \\
0 & \check{H}_{v_i v_j} \\
}, & \quad i\neq j 
\end{array}
\right.
\]
By reordering, we obtain the equivalent version
\begin{equation}\label{eq:all_Hess}
\underbrace{\mat{
H_{\delta \delta} & H_{\delta v} \\
* & H_{v v}
}
}_H := \sum_{i=1}^N \sum_{j=1}^N \Bigl( H_{ij} \otimes e_i e_j^{\sf T} \Bigr)
\end{equation}
where the submatrices are defined as
\[
\spliteq{
H_{\delta \delta} &:= {\sf diag}(\hat{H}_{\delta_i \delta_i})
,\quad
H_{\delta v} := {\sf diag}(\hat{H}_{\delta_i v_i}), \\
H_{vv} &:= {\sf diag}(\hat{H}_{v_i v_i})
+ \sum_{i=1}^N \sum_{j=1}^N \left( e_i e_j^{\sf T} \otimes \check{H}_{v_i v_j} \right).
}
\]
Using \eqref{eq:net_Hess} and the relations
\[
\spliteq{
\hat{H}_{\delta_i \delta_i} &= \gamma_i(\varrho_i;X_i)  ,\\
\hat{H}_{v_i v_i}-\hat{H}_{\delta_i v_i}^{\sf T} \hat{H}_{\delta_i \delta_i}^{-1} \hat{H}_{\delta_i v_i}
&=
\varGamma_{i}(\varrho_i;X_i) ,
}
\]
we see that \eqref{eq:cnd_iff} is equivalent to
\[
H_{\delta \delta} \succ 0
,\quad
H/H_{\delta \delta} \succeq 0.
\]
This implies that, if \eqref{eq:cnd_iff} holds, the Lyapunov function for the entire power system is positive semidefinite around the equilibrium.
Thus, the small-signal stability is proven.

\subsection{Proof of Necessity}

To prove the necessity, we utilize the fact that each submatrix of the Hessian in  \eqref{eq:all_Hess} coincides with the Jacobian obtained by linearizing the power system model around the equilibrium.
In the following, due to space limitations, we consider only the case where the two-axis generator model is connected to all buses. 
Note that the similar argument holds even when other component models are present.

Let the d-axis and q-axis symbols together be denoted as
\[
\tau_i:=(\tau_{{\rm d}i},\tau_{{\rm q}i})
,\quad
X_i':=(X_{\mathrm{d}i}', X_{\mathrm{q}i}').
\]
The linearized differential-algebraic equation system can be found as
\[
\sfdiag(I_{4N}, 0) \dot{z}^{\rm lin} = -\sfdiag(R, I_{2N}) \hat{H}' z^{\rm lin}
\]
where
\[
R :=\mat{
\omega_{0}^{-1}M^{-1} D M^{-1} & M^{-1} & 0 \\
- M^{-1}  & 0 & 0 \\
0 & 0 & \tau^{-1}(X-X') 
}
\]
with the positive definite diagonal matrices defined as
\[
\spliteq{
M&:=\sfdiag(M_i)
,\quad
\tau:=\sfdiag(\tau_i),
 \\
D&:=\sfdiag(D_i)
,\quad
X := \sfdiag(X_i)
,\quad
X' := \sfdiag(X_i').
}
\]
By eliminating the voltage phasor variables through equivalent algebraic calculations, i.e., by the Kron reduction, we obtain the ordibary differential equation system
\[
\dot{x}^{\rm lin} = \underbrace{- R (\hat{H}'/\hat{H}'_{vv})}_{A} x^{\rm lin}.
\]
Since $R+R^{\sf T}$ is positive semidefinite, the inertia theorem \cite[Corollary~4]{InertiaTheorem1962} proves that all non-zero eigenvalues of $A$ lie in the open left half-plane if and only if the Schur complement $\hat{H}'/\hat{H}'_{vv}$ is positive semidefinite.

Note that, if \eqref{eq:cnd_iff} does not hold, the Hessian in \eqref{eq:all_Hess} is not positive semidefinite at the equilibrium of interest.  
Therefore, assuming that $H_{vv}$ is positive definite, if \eqref{eq:cnd_iff} does not hold, then $H/H_{vv}$ is not positive semidefinite, proving also that $\hat{H}'/\hat{H}'_{vv}$ is not positive semidefinite from the discussion in Section \ref{sec:Hposisemi}.

Finally, let us show that $H_{vv}$ is indeed positive definite at any equilibrium.  
First, we compute the $i$th diagonal block submatrix of $H_{vv}$, denoted by $H_{v_i v_i}$.  
Noting that
\[
\check{H}_{v_i v_i} = \mat{
-Q_i -V_i^2 B_{ii} & \tfrac{P_i}{V_i} \\
* & -B_{ii}
},
\]
the sum of $\hat{H}_{v_i v_i}$ and $\check{H}_{v_i v_i}$ can be expressed as
\[
H_{v_i v_i} =
\Phi^{\sf T}_i \sfdiag (X_{{\rm q}i}^{-1},X_{{\rm d}i}^{-1}) \Phi_i
- B_{ii} \Theta^{\sf T}_i  \Theta_i
\]
where $\Phi_i$ and $\Theta_i$ are defined as
\[
\Phi_i  := \mat{
V_i \cos \phi_i & - \sin \phi_i \\
V_i \sin \phi_i & \cos \phi_i
} , \ 
\Theta_i := \mat{
- V_i \sin \theta_i & \cos \theta_i \\
V_i \cos \theta_i & \sin \theta_i 
}
\]
with $\phi_i$ in \eqref{eq:def_phi}.
Similarly, the $(i,j)$-block submatrix can be expressed as
\[
H_{v_i v_j} = - B_{ij} \Theta^{\sf T}_i  \Theta_j.
\]
From these expressions, we obtain
\[
\spliteq{
H_{vv} &= \sfdiag ( \Phi_i^{\sf T}) 
\sfdiag \Bigl(
\sfdiag (X_{{\rm q}i}^{-1},X_{{\rm d}i}^{-1}) 
\Bigr)
\sfdiag (\Phi_i )   \\
& + \sfdiag (\Theta_i^{\sf T}) (-B \otimes I_2 ) \sfdiag(\Theta_i).
}
\]
Since the first term is positive definite and the second is positive semidefinite, $H_{vv}$ is proven to be positive definite.

\section{Numerical Example} \label{sec:numerical_analysis}
\begin{figure}[t]
   \centering
   \includegraphics[width=0.6\linewidth]{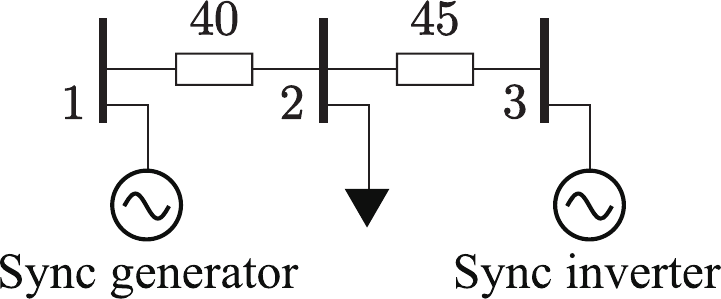}
   \caption{3-bus power system example.}
   \label{fig:simulation_setting}
\end{figure}
\subsection{Parameter Settings}
We analyze the 3-bus system in Figure \ref{fig:simulation_setting}, where Bus 1 has a synchronous generator and Bus 3 has a grid-forming inverter generator.  
We consider two cases for Bus 2: a grid-forming inverter load and a grid-following inverter load. 
Results are consistent for both VSG and FDC models.
The line susceptance is set to $40$ [pu] between Buses~1 and 2, and $45$ [pu] between Buses 2 and 3. 
The generator synchronous reactances are set to
\[
X_{{\rm d}1}=0.10,\quad
X_{{\rm q}1}=0.069.
\]
It is assumed that the grid-forming inverter generator at Bus 2 has the same reactances.  
We then vary the reactance at Bus 3 to analyze stability. 
The stationary power flow distribution of interest is shown in Table~\ref{table:power_flow}.

\begin{table}[b]
\caption{Power flow distribution.}
\label{table:power_flow}
\centering
 \begin{tabular}{lrrrr}
  \hline
& $\theta_i^{\star}$[rad] & $V_i^{\star}$[pu] & $P_i^{\star}$[pu] & $Q_i^{\star}$[pu] \\
\hline \hline
Bus~1 & $-$0.0308 & 1.0000 & 1.0000 & 0.2886\\
Bus~2 & $-$0.0560 & 0.9931 & $-$3.5000 & $-$0.5000\\
Bus~3 &  0.0000 & 1.0000 & 2.5000 & 0.3805\\
\hline
 \end{tabular}
\end{table}

\subsection{Grid-Following Inverter Load Model}
As explained in \citep{schiffer2016survey}, the standard constant power load model can be considered as a model of grid-following inverter loads that consume a certain amount of active and reactive power.  
In particular
\begin{equation}
\label{eq:load_power}
P_i = P^{\star}_i, \quad
Q_i = Q^{\star}_i
\end{equation}
regardless of the bus voltage.
Note that the active and reactive power are negative for consumption because the positive direction of power is defined as the direction from the device to the bus.

It is well known that this model also satisfies a dissipation inequality with the input $(-P_i, -\tfrac{Q_i}{V_i})$ and output $(\dot{\theta}_i,\dot{V}_i)$. 
The energy function is given by
\begin{equation} \label{eq:load_potential}
U_i (v_i)
    =
    - P^{\star}_i \theta_i - Q^{\star}_i \mathrm{ln}V_i.
\end{equation}
Note, however, that the Hessian of this energy function is always negative semidefinite if the reactive power is consumed. 
Specifically, $\varGamma_i$ in \eqref{eq:cnd_iff} for this model is
\[
\varGamma_i(\varrho_i)=
\mat{
0
&
0 \\
0 &
\frac{Q^{\star}_i}{V^2_i}
}.
\]
This implies that the grid-following inverter load disrupts the synchronization of the entire power system when it consumes reactive power.

\subsection{Results}
\begin{figure}
   \centering
   \includegraphics[width=0.5\linewidth]{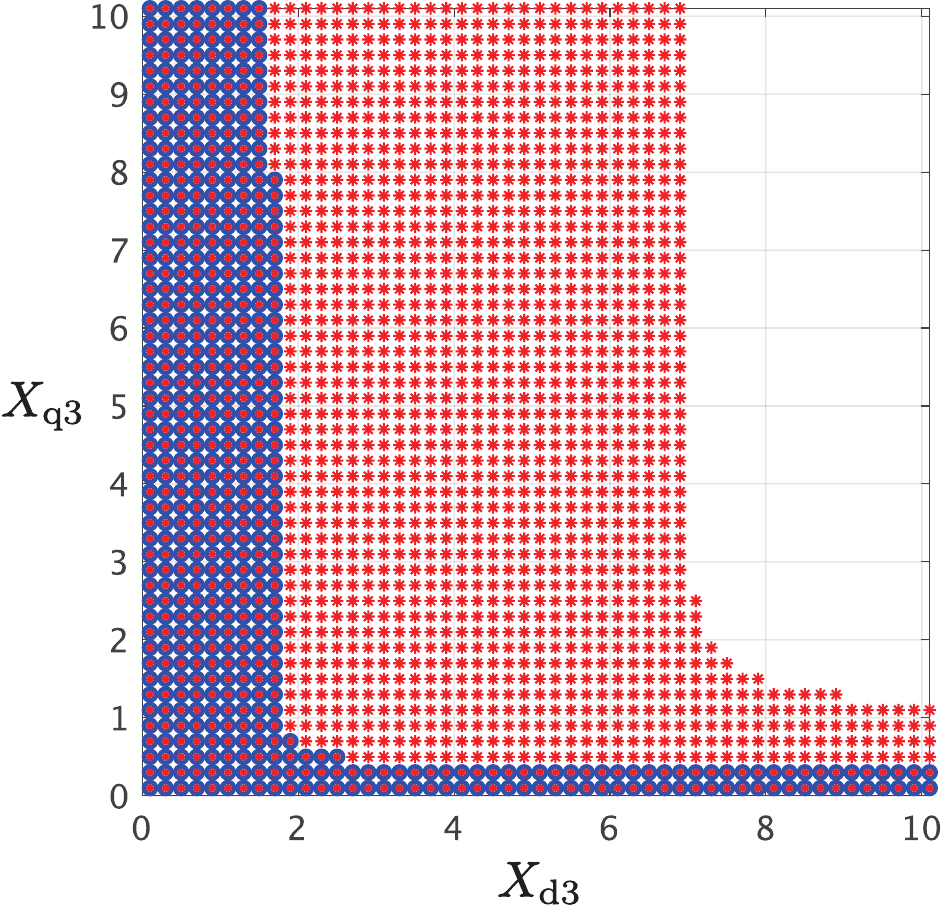}
   \caption{Stability for different synchronous reactances.}
   \label{fig:simulation_result}
\end{figure}

Let us demonstrate that, unlike the grid-following inverter load, which disrupts synchronization, a grid-forming inverter load at Bus 2 should improve it. 
Figure \ref{fig:simulation_result} shows how varying the synchronous reactance at Bus 3 affects stability. 
The red region corresponding to the grid-forming-inverter load at Bus 2 shows a wider stable range than the blue region corresponding to the grid-following inverter load.
This confirms that a grid-forming inverter improves stability compared to a grid-following inverter.


\section{Conclusion}\label{sec:conc}
This paper derived a necessary and sufficient condition for the small-signal stability of a power system with an arbitrary combination of synchronous generators and grid-forming inverters. 
The derived inequality condition is expressed as an explicit function of the solution to the algebraic equation for the stationary power flow distribution.  
Furthermore, we showed that the constant power load representing a grid-following inverter load acts as a device that disrupts the synchronization of the power system. 
Future work includes extending the theory to cases with transmission losses and considering more detailed internal dynamics of inverters.

\begin{ack}
This paper is based on results obtained from a project, JPNP24007, commissioned by the New Energy and Industrial Technology Development Organization (NEDO).
\end{ack}

\bibliography{bib/citation,bib/reference,bib/reference_CREST} 
\appendix
\end{document}